\documentclass[conference, 10pt, a4paper]{IEEEtran}
\IEEEoverridecommandlockouts

\usepackage[left=0.65in, right=0.65in, bottom=1in, top=0.75in]{geometry}
\usepackage{cite}
\usepackage{amsmath,amssymb,amsfonts}
\usepackage{algorithmic}
\usepackage{graphicx}
\usepackage{textcomp}
\usepackage[table,xcdraw]{xcolor}
\usepackage{romanbar}
\usepackage{tikz}
\usepackage{mathdots}
\usepackage{yhmath}
\usepackage{cancel}
\usepackage{color}
\usepackage{siunitx}
\usepackage{array}
\usepackage{multirow}
\usepackage{gensymb}
\usepackage{tabularx}
\usepackage{booktabs}
\usepackage{verbatim}
\usepackage[ruled,vlined]{algorithm2e}
\usepackage{balance}

\usetikzlibrary{fadings}
\def\BibTeX{{\rm B\kern-.05em{\sc i\kern-.025em b}\kern-.08em
		T\kern-.1667em\lower.7ex\hbox{E}\kern-.125emX}}
\pdfpagewidth=\paperwidth
\pdfpageheight=\paperheight
\begin{document}

\title{Performance Evaluation of Transmission Mode Selection in D2D communication
\thanks{This research is part of a project that has received funding from the European Union's Horizon 2020 research and innovation programme under grant agreement Nº739578 and the government of the Republic of Cyprus through the Directorate General for European Programmes, Coordination and Development.}
}

\author{\IEEEauthorblockN{Iacovos Ioannou\authorrefmark{1}\authorrefmark{2}, Christophoros Christophorou\authorrefmark{1}\authorrefmark{3}, Vasos Vassiliou\authorrefmark{1}\authorrefmark{2}, and Andreas Pitsillides\authorrefmark{1}}
\IEEEauthorblockA{\authorrefmark{1}Department of Computer Science, University of Cyprus\\
\authorrefmark{2}RISE Center of Excellence on Interactive Media, Smart Systems and Emerging Technologies, Nicosia, Cyprus \\
\authorrefmark{3}CITARD Services LTD, Nicosia Cyprus \\
}
}

\maketitle

\begin{abstract}
Device to Device (D2D) Communication is expected to be a core part of the forthcoming 5G Mobile Communication Networks as it promises improvements in energy efficiency, spectral efficiency, overall system capacity, and higher data rates with the use of the same frequencies for different D2D transmissions in short communication distances within the Cell. However, in order to achieve optimum results, it is important, among others, to select wisely the Transmission Mode of the D2D Device. Towards this end, our previous work proposed an intelligent Transmission mode selection approach in a framework that is utilizing Artificial Intelligence (AI) BDIx agents to collectively satisfy the D2D challenges in a Distributed Artificial Intelligent (DAI) manner autonomously and independently. In this paper, as a first step, a literature review focused on related Transmission mode approaches, is performed. Then, our investigated Transmission mode selection approach is further explained with formulas and evaluated based on different threshold values and investigated how these can affect the overall spectral efficiency and power usage of the network in order to achieve the maximum performance. The investigated thresholds(i.e. D2D Device Weighted Data Rate (WDR) and the D2D Device Battery Power Level) and metrics(i.e. WDR) are also further analyzed and formulated. In addition, the effect the transmission power of the D2D links has on the total spectral efficiency and total power consumption of the network, is also examined.  This evaluation results arise some interesting findings that can contribute in other approaches that utilized similar or same thresholds. Also, the results obtained demonstrate that with the right tuning of the thresholds and transmission power, one can achieve a significant improvement in the network power usage and total spectral efficiency. 
\end{abstract}

\begin{IEEEkeywords}
5G, D2D, Transmission Mode Selection, Distributed Artificial Intelligence
\end{IEEEkeywords}

\section{Introduction}
\label{intro}

Device to Device (D2D) communication is expected to be a core part of the forthcoming 5G Mobile communication network. The main reason is that, in contrast with cellular communications, D2D communications are not constrained by the licensed frequency bands; i.e., the full radio spectrum, both licensed (inband D2D) and unlicensed (outband D2D), can be utilized. In addition, D2D Communication is transparent to the cellular network as it permits adjacent User Equipment (UEs) to bypass the Base Station (BS) and establish direct links between them, either by sharing their connection bandwidth and operate as relay stations, or by directly communicating and exchanging information. For the aforesaid reason, D2D can be used to implement numerous of the 5G requirements and consequently improve spectral efficiency, data rates, throughput, energy efficiency, delay, interference and fairness   \cite{Doppler2009,Fodor2012,Gandotra2016,Ahmad2017}. In addition, due to the shorter communication distance, D2D can offer lower power consumption for the communicating D2D devices and more prominent sum rate due to D2D cluster support (i.e. traffic offloading) \cite{Gandotra2016,Ahmad2017}. However, in order to actualize D2D in 5G, several challenges need to be addressed, including Device Discovery, Mode Selection, Interference Management, Power Control, Security, Radio Resource Allocation, Cell Densification \& Offloading, QoS \& Path Selection, use of mmWave Communication, Non-cooperative users, and Handover Management \cite{Ioannou2020}. The D2D devices can act as network elements in the 5G architecture with the characteristic to choose frequencies and share bandwidth. By utilizing D2D approaches, a distributed ultra-dense network can be created \cite{ultradense} under the mobile cellular network. However, in order to achieve optimum results, it is important, among others, to select wisely the Transmission Mode (i.e., D2D Relay, D2D Multihop Relay, D2D Cluster) of the D2D Device. The main reason is that the Transmission Mode selected for a device can affect the creation of the clusters, the way data will be communicated between the D2D Devices, and optimize backhauling links between disconnected/disjointed clusters by forming better paths. Towards this end, in previous work \cite{Ioannou2020} we proposed an intelligent Transmission Mode Selection algorithm called DAIS, as part of a framework that is utilizing Artificial Intelligence (AI) agents to collectively satisfy the D2D challenges in a distributed manner with the use of BDIx agents.

This paper contains a more focused literature review in terms of D2D Transmission Mode selection and a comprehensive evaluation of the DAIS Transmission Mode Selection algorithm, in a challenging and dynamic environment such as D2D communication.  Different parameters affecting the network's spectral efficiency and power usage have been determined,analyzed and formulated. The parameters that are tuned in this work are the Device Battery Threshold and the Weighted Data Rate Threshold. The results obtained demonstrate that with the right tuning of the thresholds we can have a significant improvement in the network power usage and total spectral efficiency. In addition, we have also examined how the maximum transmission power of each device can affect the total spectral efficiency and power consumption in the network.

The rest of the paper is structured as follows. Section II provides background information and related work associated with Transmission Mode Selection. Section III presents the problem formulation of the examined approach. Specifically, we show the implementation of the DAIS algorithm, outlining the assumptions, constraints, thresholds and new metrics that are introduced. The performance of the proposed Transmission Mode Selection approach is examined and evaluated in Section IV, Finally, section V contains our Conclusions and Future Work.

\section{Background Knowledge and Related Work}

\subsection{Background Knowledge}\label{CD2D}
This section provides background knowledge regarding the main characteristics of D2D communications. We explain the types of Control that can be exploited for the establishment of D2D Communication links, as well as the types of Transmission modes that a D2D Device can operate as.

\subsubsection{Types of Control in D2D Communication}\label{TypesOfControl}
We can categorize the types of control that can be used for the establishment of D2D Communication links, as follows:

\begin{itemize}
	\item Centralized: In the centralized technique, the BS completely oversees the UE nodes, even when they (UEs) are communicating directly. The controller oversees interference/connections/path etc., amid cell with D2D UEs.
	\item Distributed: Within a distributed scheme, the procedure of D2D node (interference/data rate/path) management does not oblige to a central entity, but it is performed autonomously by the UEs themselves. The distributed scheme diminishes the control and computational overhead. This scheme is particularly appropriate for large size D2D networks. In a distributed system all control processes are expected to begin at simultaneously and run in parallel.
	\item Semi distributed: Both centralized and distributed schemes have strong points and drawbacks. Trade offs can be accomplished between them for better performance. Such D2D management schemes are referred to as "semi-distributed" or "hybrid".
\end{itemize}

\subsubsection{Types of Transmission Modes in D2D Communication}
\begin{figure}
	\centering
	\includegraphics[width=1\linewidth]{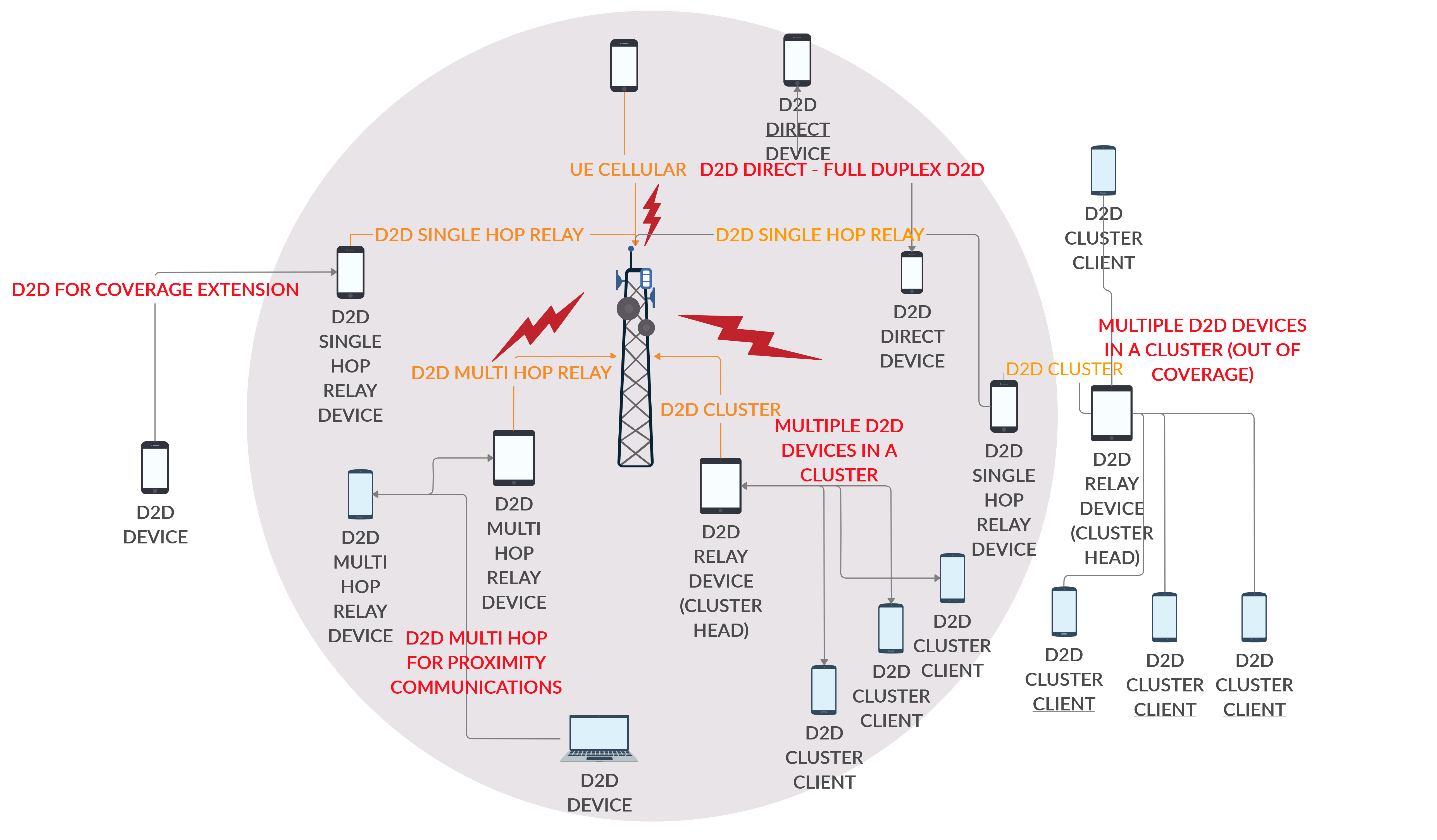}
	\caption{Types of Transmission Modes in D2D Communication}
	\label{fig:typesoftrasmissionmode}
\end{figure}

There exist different transmission modes for D2D communication (see Fig. \ref{fig:typesoftrasmissionmode}), based on how UEs interact with the BS and other D2D nodes shown in \cite{Ioannou2020}.
\begin{itemize}
	\item D2D Direct: Two UEs connect to each other by utilizing licensed or unlicensed spectrum. The two D2D UEs only communicate among each other.
	\item D2D Single-hop Relaying: Contribution of bandwidth between a UE and other UEs. In D2D Single-hop Relaying mode one of the D2D UEs is connected to a BS or Access Point and provides access to an additional D2D UE. \cite{Deng2015a}.
	\item D2D Multi-hop Relay: The single-hop mode is extended by empowering the connection of more D2D UEs in chain. Both backhaul and D2D transmissions are performed in an uplink with other D2D relay node (as a bridge) and they are subject matter to the control of the former D2D relay node \cite{Steri2016}.		
	\item D2D Cluster: D2D cluster is a group of UEs connected to a D2D relay node performing as a Cluster Head (CH). The D2D relay node acts as an intermediate router to the network through an access point or BS. Clustering is appropriate in high user densities \cite{Song2014,Peng2013}.
\end{itemize}

\subsection{Related Work on Transmission Mode Selection in D2D Communication}
\label{RTS}

Approaches related to the Transmission mode selection investigated in this paper, are provided in a plethora of articles \cite{Doppler2010,Jung2012,Pang2013,Han2012,Xiang2012,Liu2016,Xu2017,Ma2012,Ma2016,Zhao2016,Feng2015,Wang2013b,Kim2014a,Kazmi2017}. Below we refer only to those that are most relevant to the work investigated in this paper.

\begin{figure}
	\centering
	\includegraphics[width=1\linewidth]{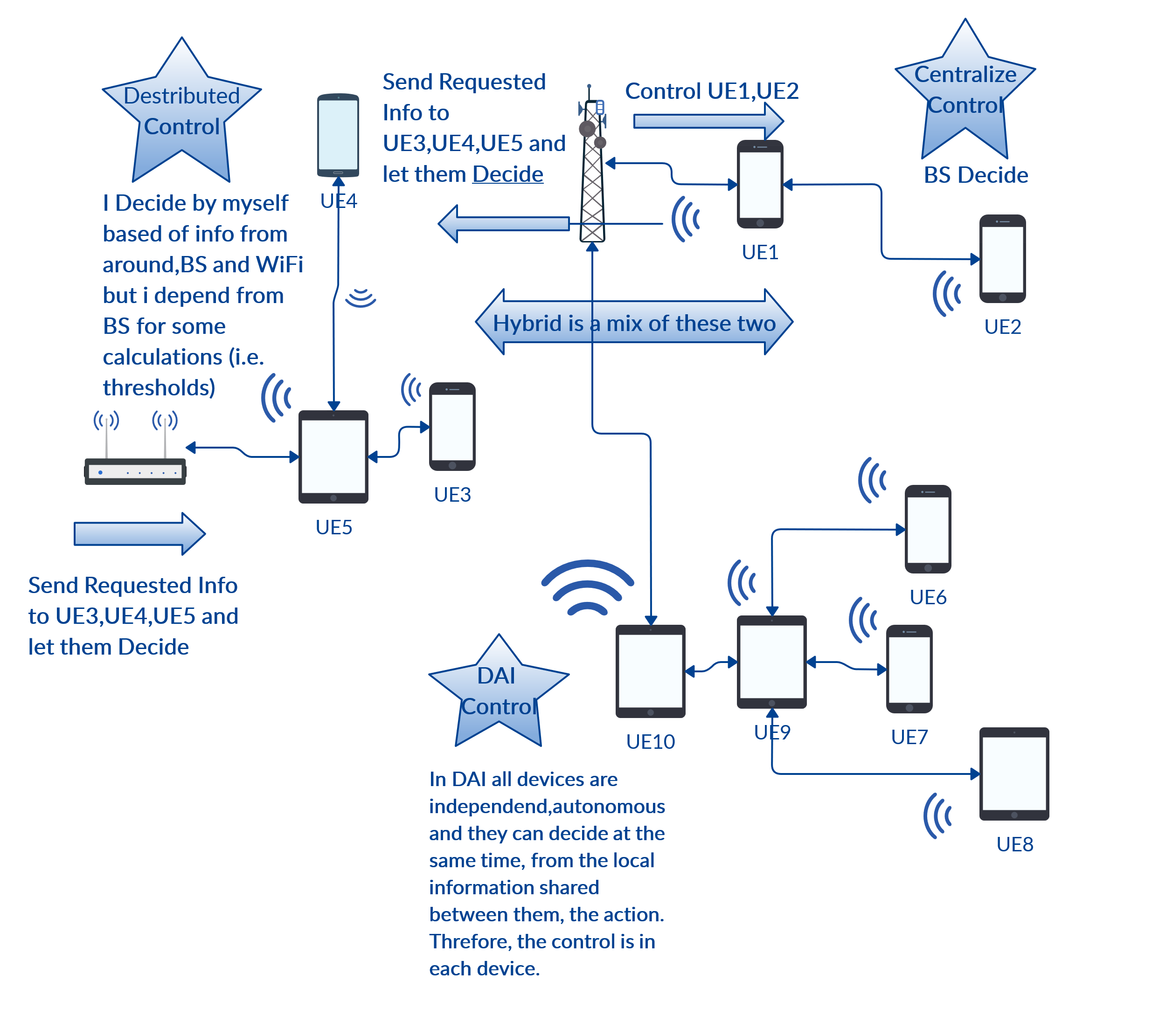}
	\caption{Types of Control in D2D Communication}
	\label{fig:types-of-control}
\end{figure}

A classification based on the type of control (see Section \ref{TypesOfControl} and Fig. \ref{fig:types-of-control}) used by each paper examined is found below:
\begin{itemize}
	\item Centralized approaches \cite{Doppler2010,Jung2012,Pang2013,Han2012,Xiang2012,Xu2017,Ma2016,Zhao2016,Feng2015,Wang2013b,Kim2014a,Rigazzi2014,Gui2018}, where the decision is taken by the BS;
	\item Semi-distributed approaches \cite{Liu2016}, where the decision is taken by both the BS and the D2D Devices in collaboration;
	\item Distributed approaches \cite{Ma2012}, where the decision is taken by the D2D Devices; however in this case the D2D Devices need some information from the BS; and
	\item Distributed Artificial Intelligent (DAI) approaches, where the decision is taken by each D2D Device independently; however in this case they may share information with other D2D Devices  \cite{Ioannou2020}.
\end{itemize}

It is evident from this preliminary analysis that most works use the Centralized approach and only a few use Semi- or Fully-Distributed algorithms.

The metrics considered for selecting the Transmission mode to be adopted are shown in Table \ref{Table:metrics}. There is, again, a concentration of works using the same metrics (Power, SINR, Distance).

\begin{table}[]
	\centering
	\caption{Metrics Utilized in D2D Transmission Mode Selection}
	\label{Table:metrics}	
	\begin{tabular}{|c|c|}
		\hline
		 \textbf{Metrics} &  \textbf{Works using the metric} \\ \hline
		Power or Transmission Power	& \cite{Xiang2012,Xu2017,Ma2012,Ma2016,Kim2014a}                 \\ \hline
		Interference & \cite{Ma2012,Wang2013b}                     \\ \hline
		Resource Blocks or Sub-channel	&  \cite{Pang2013}       \\ \hline
		SINR & \cite{Doppler2010,Jung2012,Pang2013,Liu2016,Xu2017,Zhao2016}         \\ \hline
		Channel Signal Indicator (CSI) & \cite{Rigazzi2014}        \\ \hline
		Distance  & \cite{Xiang2012,Feng2015,Gui2018,Ioannou2020}                  \\ \hline
		Hop Count (in Multi Hop Relays) & \cite{Rigazzi2014}       \\ \hline
		Sum Rate or Type of frequencies &  \cite{Han2012,Wang2013b}  \\ \hline	
		Battery Capacity &  \cite{Gui2018}  					\\ \hline		
		Data Forwarding Delay &  \cite{Gui2018}  \\ \hline		
		Link Throughput to eNB (BS) &  \cite{Gui2018}  \\ \hline	
		Weighted Data Rate &  \cite{Ioannou2020}  \\ \hline													
	\end{tabular}
\end{table}

The following approaches focus on D2D Device Selection only \cite{Doppler2010,Jung2012,Pang2013,Han2012,Liu2016}. More specifically, in \cite{Doppler2010} the authors use only the quality of the cellular link and interference (SINR) and a simple condition to select the best D2D Device to connect.  In \cite{Jung2012} the authors are also using the SINR, but with the target to maximize the sum rate by using a gradient method. In \cite{Pang2013} the authors, in addition to SINR, consider Sum Rate as well, by utilizing an evolutionary algorithm. In \cite{Han2012} the aim is to maximize the average Sum Rate by utilizing an opportunistic subchannel scheduling to solve a stochastic optimization problem. The authors in \cite{Liu2016} use SINR and Lagrangian dual decomposition method in conjunction with a greedy and a column generation based algorithm. With this approach a threshold calculation is first executed at the BS (Lagrange multiplier). Then, the UEs based on the calculated threshold perform a decision independently.

There are approaches that focus on D2D Direct and D2D Relay selection mode only \cite{Xiang2012}, \cite{Xu2017,Ma2012,Ma2016,Zhao2016,Feng2015,Wang2013b,Kim2014a}. More specifically, in \cite{Xiang2012} the authors use the power usage as a metric, and propose a distance-dependent algorithm with power optimization based on the UE position. In \cite{Xu2017}, using as utility the power and the SINR, the authors select the best D2D Relay by tackling a mixed integer nonlinear programming problem using both a two-dimensional and a three-dimensional matching. In \cite{Ma2012} the authors choose a D2D Relay by utilizing interference as a metric. In this approach a distributed method is chosen to coordinate the interference and eliminate improper D2D Relays by minimizing power. In \cite{Ma2016} the authors formulate the D2D Relay selection problem as a combination optimization many-to-one matching problem. Power is used as a metric  in their Power efficient Relay Selection algorithm. In \cite{Zhao2016}, by using SINR as a metric, a two-stage D2D Relay selection is proposed. In the first stage, the range of the candidate D2D Relay UEs are determined by using a regional division method. In the second stage the optimal D2D Relay UE is selected. In \cite{Feng2015}, by using distance as a metric, a multi-cell model based on stochastic geometry is proposed. The aim of this model is to evaluate the coverage probability of three location-aware relay selection schemes. In \cite{Wang2013b} the authors based on outage probabilities analysis and a sum-capacity comparison provide the criteria of employing Relay communication mode with two hops. The metric used in this analysis is interference that is calculated based on Sum Rate. In \cite{Kim2014a}, by using power as a metric, an iterative Hungarian method (IHM) is proposed to solve the optimal power allocation problem. This method takes under consideration the channel allocation.

Furthermore, there are approaches that focus on D2D Direct and D2D Multi hop Relay \cite{Rigazzi2014,Gui2018}. More specifically, in \cite{Rigazzi2014}, the authors are using graph theory (Destination Oriented Directed Acyclic Graph (DODAG)) to provide, by means of multi-hop path, the location of D2D nodes in the cluster network topology. Initially, by using as a metric the CSI, the BS concludes with the potential D2D multi hop relays and D2D Devices. Then, the hop count metric is utilized as  a cumulative cost function to construct the graph. In \cite{Gui2018}, the authors propose an Ordinal Potential Game (OPG), with the purpose to select the best link and association between D2D nodes. In this approach, the transmission mode selection is performed as a throughput maximization problem with delay and remaining energy constraints. The metrics used for the selection are the location information, battery capacity, data forwarding delay, and the link throughput associated with it to the eNB (BS).

It is worth mentioning that all the approaches investigated above, separate the UEs into categories. Those that are candidate to become D2D Devices and those that will stay connected to the BS as regular UEs. On the other hand, our approach \cite{Ioannou2020} considers all the UEs as candidates to become a D2D Device, which can provide better network performance. Furthermore, to the best of our knowledge, there is not any other approach in the open literature that tackles the problem of having a D2D Device utilizing all transmission modes (D2D Relay, D2D multi-hop and D2D cluster) in a distributed manner. Also, the investigated approach \cite{Ioannou2020} with the Distance as metric, it is utilizing a new introduced metric, the Weighted Data Rate (WDR). With the new metric the investigated paper achieves transmission mode selection, in a distributed manner.

\section{Problem Formulation at Distributed Artificial Intelligence Solution (DAIS) Algorithm for Transmission Mode Selection}
\label{section:problemFormulation}
The problem that this paper tries to tackle is to maximize the total spectral efficiency (sum rate) and reduce the power consumption of the DAIS algorithm. More specifically, from Shannon–Hartley theorem, the spectral efficiency is shown in Equation \ref{SpectralEfficiency}, measured in (bits/s/Hz), where ''C is capacity'' (in bits per second), ''B is bandwidth'' (hertz) , ''S is signal power'' and ''N is noise power'' (in decibel dB). Moreover, for the mobile and wireless networks with a single-antenna and point-to-point scenario, the AWGN spectral efficiency from channel capacity is used with the power-limited and bandwidth-limited scheme as indicated in Equation \ref{AWGNspectralefficiency}, measured in  (bits/s/Hz), where the average received power (in W) is calculated using Free Space Model and Free Space Path Loss as $\bar{P}$, Transmission Power is known to the channel (it is the TP), Power Consumption is shown in Equation \ref{PowerConsumption}, SNR is the received signal-to-noise ratio (SNR) and lastly the noise is  $N_0$ (W/Hz).Therefore, the problem is based on the Equations \ref{totalSpectralEfficiency} and \ref{powerConcuption} with the purpose to find the best WDR Threshold and Device Battery Threshold in order to achieve the maximization/minimization. This is a NP-hard problem to solve. 

\begin{equation}
\label{SpectralEfficiency}
SE=\frac{C}{B} = \log_2 \left( 1+\frac{S}{N} \right)\
\end{equation}

\begin{equation} \label{AWGNspectralefficiency}
\begin{split}
SE & = \frac{C_{\text{AWGN}}}{W}=\log_2\left(1+SNR\right)  \\
SNR & = \frac{\bar{P}}{N_0 W}
\end{split}
\end{equation}

\begin{equation}
\label{PowerConsumption}
PC=TP-\bar{P}
\end{equation}

\begin{equation}
\label{totalSpectralEfficiency}
Total SE=\max {\sum\limits_{i=1}^{\#UEs}}{SE}
\end{equation}

\begin{equation}
\label{powerConcuption}
Total PC=\min {\sum\limits_{j=1}^{\#UEs}}{PC}
\end{equation}

In this section we show the implementation of the investigated intelligent approach shown in \cite{Ioannou2020}. The investigated approach calculates Data Rate, total Spectral Efficiency (Sum  Rate) and power consumption as shown in \cite{Zhou2014a}. The investigated approach introduced a new metric, called ”Weighted Data Rate” (WDR). The WDR is defined at each node in D2D communication as the minimum Data Rate in the path that the D2D Device selected, whether it is directly connected to the BS or through another D2D Device. The aim of the approach is to maximize the WDR, i.e., WDR = max(min(Link Data Rate) for each path. More specifically, at the beginning, the entering D2D Device has as WDR ($WDR_0$) the data rate of the link to BS (using eq. \ref{eq:1}). Afterwards, it selects the most optimum path from each Relay (D2DRelay/D2DMultiHopRelay) and its Transmission mode with the use of DAIS algorithm (shown in \cite{Ioannou2020} by using eq. \ref{eq:2} and \ref{eq:3}). Then, it calculates its own WDR (using eq. \ref{eq:4}). Note that, that WDR of Relay is the same of WDR\_Path and the WDR is shared as a message advertisement with the use of LTE proximity services. More details about the problem formulation, the assumptions, constrains, sum rate and power consumption estimations (utilized from \cite{Zhou2014a}), parameters, thresholds calculations and terminology as well as our investigated Transmission Mode selection algorithm can be found in \cite{Ioannou2020}.

{\tiny
\begin{align}
	\label{eq:1}
	WDR_0(D2D)=LinkDataRate(D2D,BaseStation) \\
	\label{eq:2}
	MinPath(x)=\min_{y=U_1, \dots, U_N 	\in Path(y) }(LinkDataRate(y,y+1))\\
	\label{eq:3}            
	WDR_Path=\max_{x=Relay, \dots, RelayN}(MinPath(x))\\ 
	\label{eq:4}
	WDR(D2D)=\min(LinkDataRate(D2D,WDR_Path),WDR_Path)
\end{align}
}%

\section{Performance Evaluation}
When DAIS was first presented it was accompanied by a simple proof of concept evaluation. In this work we investigate its sensitivity and performance while varying the number of UEs in the network, the Device Battery Threshold, and the Weighted Data Rate threshold. In addition we examine how transmission power change the aforesaid metrics.

\subsection{Methodology}

\begin{itemize}
	\item The Device Battery Threshold: This threshold determines the minimum value (in \%) that a D2D Device must have in remaining battery, in order to become D2D Relay or D2D multi hop Relay and accept connections from other UEs; and
	\item The Weighted Data Rate (WDR) threshold: This threshold determines: i) the minimum WDR that an existing D2D Device operating as D2D Relay/ D2D multi hop Relay must have in order for a new D2D Device entering the network to connect to it; or ii) the maximum WDR that a new D2D Device entering the D2D Network must have in order replace a D2D Device operating as D2D Relay/ D2D multi hop Relay and take its role.
\end{itemize}

The Battery Power Threshold is used by the algorithm in order to conserve energy for D2D Devices acting as D2D Relay or D2D Multi Hop Relay Devices. More specifically, a D2D Relay or D2D Multi Hop Relay Device will admit connections to new D2D Devices entering the Network only when their battery power level is greater or equal than the threshold.

The WDR threshold is used by the algorithm for four purposes. Through the WDR Threshold, an entering new D2D Device in the Network:

\begin{itemize}
	\item Can perform a quality check of the D2D Relay, in order to connect to it as a D2D Client. The new D2D Device entering the network, will: \textbf{i)} at first, find all the D2D Relays in its proximity and sort them based on their WDR; \textbf{ii)} as a second step, filter them using the WDR based on the eq. \ref{eq:5};	and \textbf{iii)} in the last step select and connect to the best D2D Relay with the highest WDR.
	\item Can perform a quality check of the D2D Multihop Relay, in order to connect to it either as a D2D Client or a D2D Relay (this is based on the distance of the D2D Device from the D2D Multihop Relay Device; for more information see Section \ref{section:problemFormulation}). The steps followed are the same as above.
	\item Can replace a D2D Relay/ D2D multi hop Relay Device and take its role, if the new D2D Device WDR is greater than the WDR of the existing D2D Relay/D2D multi hop Relay Device. The new D2D Device entering the network, will: \textbf{i)} in the first step find all the D2D Relay/D2D multi hop Relay Devices in its proximity and sort them based on their WDR; \textbf{ii)} in the second step filter them using the WDR based on the eq. \ref{eq:6}; and \textbf{iii)} in the last step select and replace the D2D Relay D2D multi hop Relay Device with the worst WDR.
	\item Can connect to a D2D Relay/ D2D multi hop Relay Device in its proximity, and act as a D2D Relay. The new D2D Device entering the network, will: \textbf{i)} in the first step find all the D2D Relay/ D2D multi hop Relay Device with no connection in its proximity and sort them based on their WDR; \textbf{ii)} in the second step filter them using the WDR based on the eq. \ref{eq:7}; and \textbf{iii)} in the last step select and connect as D2D Relay to the best a D2D Relay/ D2D multi hop Relay Device with the highest WDR. In case the connection is made with a D2D Relay Device, the D2D Relay Device will change to D2D Multihop Relay.
\end{itemize}

{\tiny
	\begin{align}
	\label{eq:5}
	(WDR_{Threshold}+1) \times WDR_0(D2D) \leqslant D2DR_{WDR} \\
	\label{eq:6}
	D2DR_{WDR} \lor D2DMHR_{WDR} \leqslant (WDR_{Threshold}-1) \times WDR_0(D2D) \\
	\label{eq:7}
	D2DR_{WDR} \lor D2DMHR_{WDR} \geqslant (WDR_{Threshold}+1) \times WDR_0(D2D) 
	\end{align}
}%

\subsection{Simulation Environment}
In our system model we used a variable number of D2D Devices in order to evaluate the results and achieve the best thresholds of the algorithm, while using the same constraints and simulation parameters as in \cite{ Ioannou2020}. In the simulation environment we utilize Matlab "LTE Toolbox" with an implementation in JAVA of the calculation of transmission modes of each D2D Device based on the WDR. More specifically, 10 to 1000 D2D Devices are placed in a cell range of 1000 meter radius using the Poisson Point Process (PPP) distribution, with the BS located at the center of the cell. The battery power level of the D2D Devices is determined by using a probability estimation function following Gaussian distribution of mean 0.70 and standard deviation 0.30. Also, the Sum Rate of the D2D network is calculated with the same way as in \cite{ Ioannou2020}, basically by adding all the achieved data rates of all nodes in the network. Both thresholds are examined for a range of values from 0\% until 100\% with a step of 0.05\%.

The transmission power of the D2D communication link by each device was also examined. We reduced the transmission power from 160 mWatt down to 60 mWatt with a step of -10 mWatt. For the estimation of the changes occurred in the spectral efficiency and power consumption results, the optimal WDR and Battery Power Threshold values extracted in the Results section (Section \ref{results}) have been considered in the simulation. The aim of this investigation is to examine and prove that the clusters created by our algorithm using the WDR as a metric, are in the best positions.

\subsection{Results}
 \label{results}
\begin{figure}[!tbp]
	\centering
	\begin{minipage}[b]{0.40\textwidth}		
		\includegraphics[width=\linewidth]{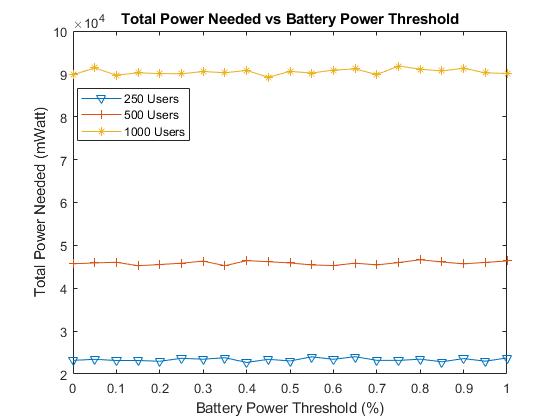}
		\caption{Different Number of Devices with Power Needed Vs Power Threshold}
		\label{fig:DifferentNumberofDevicesandPowerNeededvsPowerThreshold}		
		
	\end{minipage}
	\hfill
	\begin{minipage}[b]{0.40\textwidth}
		\includegraphics[width=\linewidth]{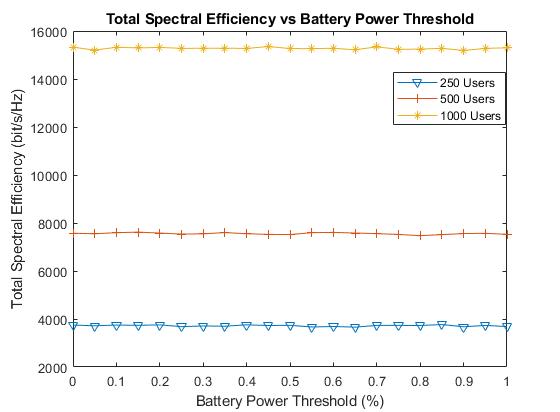}
		\caption{Different Number of Devices with Sum Rate Vs Power Threshold}
		\label{fig:sum-rate-vs-power-threshold}
	\end{minipage}
\end{figure}

\begin{figure}[!tbp]
	\centering
	\begin{minipage}[b]{0.40\textwidth}		
		\includegraphics[width=\linewidth]{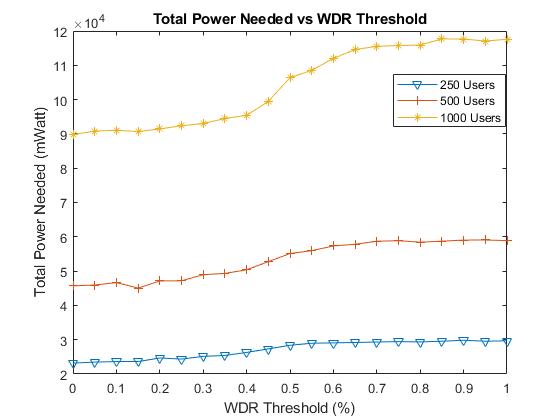}
		\caption{Different Number of Devices with Power Needed Vs WDR Threshold}
		\label{fig:power-needed-vs-wdr-threshold}	
	\end{minipage}
	\hfill
	\begin{minipage}[b]{0.40\textwidth}
		\includegraphics[width=\linewidth]{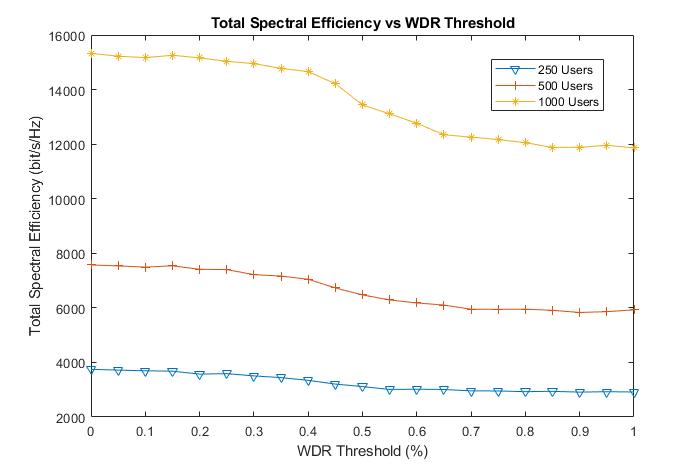}
		\caption{Different Number of Devices with Sum Rate Vs WDR Threshold}
		\label{fig:sum-rate-vs-wdr-threshold}
	\end{minipage}
\end{figure}

A brute force investigation for finding the optimum thresholds is executed. The investigation of the approach’s spectral efficiency is shown in figures \ref{fig:sum-rate-vs-power-threshold} and \ref{fig:sum-rate-vs-wdr-threshold}. Also the examination of power needed for establishing the D2D communications (i.e., power reserved by utilizing both thresholds) is shown in figures \ref{fig:DifferentNumberofDevicesandPowerNeededvsPowerThreshold} and \ref{fig:power-needed-vs-wdr-threshold}.

\begin{figure}[!tbp]
	\centering
	\begin{minipage}[b]{0.40\textwidth}
		\includegraphics[width=\linewidth]{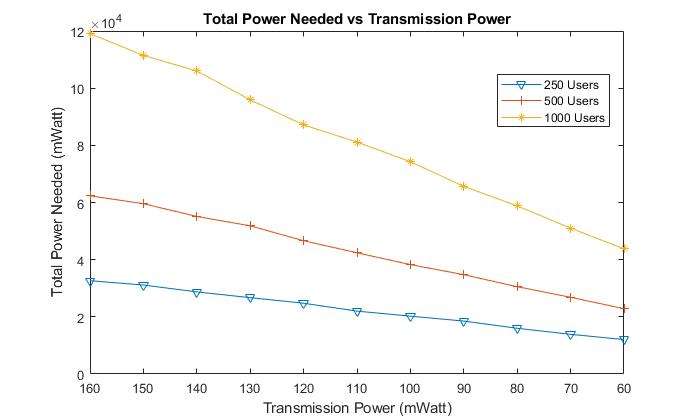}
		\caption{Transmission Power of Communication Vs Total Power needed for Communication}
		\label{fig:powerForPowerNeeded}
	\end{minipage}	
	\begin{minipage}[b]{0.40\textwidth}
		\includegraphics[width=\linewidth]{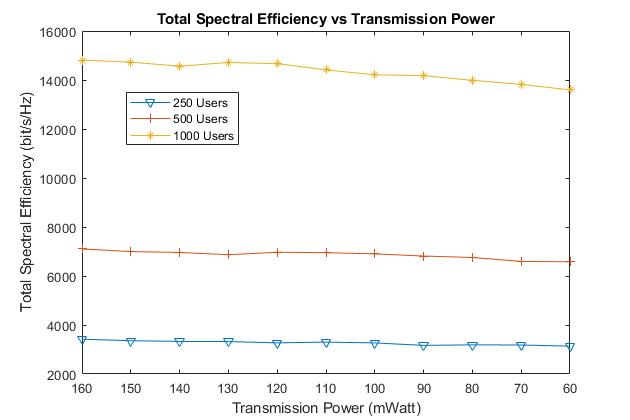}
		\caption{Transmission Power of Communication Vs Total Efficiency}
		\label{fig:powerForSumarate}
	\end{minipage}
	\hfill
\end{figure}

As observed from the results, changes in the battery power threshold does not cause any changes in the power usage and total spectrum efficiency. More specifically, as shown in Figure \ref{fig:DifferentNumberofDevicesandPowerNeededvsPowerThreshold} and Figure \ref{fig:sum-rate-vs-power-threshold}, there are not any major changes in the power consumption or spectral efficiency with a different number of users and different battery power thresholds. The reason is that the battery power level does not affect the formation of the D2D communication network. This is because the battery power of the D2D Device is calculated using a Gaussian distribution with mean 0.60 and standard deviation 0.40.

On the other hand, altering the WDR Threshold (called PERCDATARATE in the algorithm) affects significantly the results, improving the efficiency and reducing power consumption. More specifically, as shown in Figure \ref{fig:power-needed-vs-wdr-threshold} and Figure \ref{fig:sum-rate-vs-wdr-threshold}, with a different number of users and different WDR threshold there are major changes in the resulting power consumption and in total spectral efficiency.

Overall, as depicted in the above figures, the WDR threshold achieving optimized results is 35\%. On the other hand, even if the battery power threshold is not affecting significantly the spectrum efficiency and power consumption, best results can be achieved with a threshold value above 75\%. It is important to note here that the battery power threshold does not alter statistically the power consumption nor the sum rate (total spectral efficiency) because of the Gaussian distribution that the D2D Devices’ battery power follow. More specifically, due to the low number D2D Relay candidates to be selected as cluster heads (D2D Relays), the best selection is succeeded independently of the battery power threshold.

By altering the transmission power of the communication we observed gains on the power consumption and the spectral efficiency. More specifically, as shown in figure \ref{fig:powerForPowerNeeded}, demonstrating the power consumption for the D2D communications, when the transmission power used for all D2D communication links is decreased, the total power consumption decreases drastically with a maximum change of 63.23\%. However, as shown in figure \ref{fig:powerForSumarate}, there is no reciprocal change for the total spectral efficiency, as this value changes with a smaller rate reaching a maximum degradation of 8.22\%.

Thus, as observed above, the gain of decreasing transmission power is major both for the conservation of energy and the maximization of spectral efficiency. More specifically, by reducing the transmission power of the D2D communication links, the Efficiency of the network is not actively affected. On the other hand, as shown in Figure \ref{fig:powerForPowerNeeded} and in Figure \ref{fig:powerForSumarate}, the power consumption is drastically affected if more D2D Devices are introduced and form clusters in the Network.

Additionally, a preliminary comparison of our Transmission mode selection algorithm \cite{Ioannou2020} with another related approach \cite{Ma2016} have been performed, demonstrating significant benefits on the average power consumption of a D2D Device. More specifically, the results provided in \cite{Ma2016} shows that algorithm creates 10 D2D Relays and 20 D2D communication links. Instead, in our algorithm, because cluster creation depends on the number of the D2D Devices that exist in the network, the algorithm creates a maximum number of 12 D2D Relay Devices and a maximum number of 1988 D2D communication links. Therefore, comparing our approach with the one described in \cite{Ma2016} using the same scenario, we observe significant gains on the average power consumption of the network. In particular, in \cite{Ma2016}, the average power consumption for a D2D  was  0.23 W (for 20 UEs). On the other hand when our approach was simulated this was reduced to 81mW (for 1000 UEs).  Also in contrast to \cite{Ma2016}, our approach was not restricting the data rate.

\section{Conclusions}
In this paper, the performance of the DAIS Transmission Mode Selection approach was examined and evaluated. More specifically, different threshold values, related to D2D WDR and Battery power level, have been used affecting spectral efficiency and power usage of the network and those achieving optimum performance have been determined. The results obtained demonstrate that with the right tuning of the thresholds we can have a significant improvement in the network power usage and total spectral efficiency. Overall, from the results collected, a WDR threshold of 35\% and a battery power threshold of 75\%, provides the best results. In addition, it was examined how the total spectral efficiency and power reservation are affected by altering the transmission power of the communication. More specifically, when the transmission power used for all D2D communication links is decreased, the total power consumption decreases drastically with a maximum change of 63.23\%. However, this is not replicated for the total spectral efficiency as well, as the latter changes with a smaller rate reaching a maximum degradation of 8.22\%.

\balance

	\bibliography{conference_ieee}
	\bibliographystyle{ieeetr}

\end{document}